\documentclass[conference]{IEEEtran}
\IEEEoverridecommandlockouts
\usepackage{booktabs}
\usepackage{cite}
\usepackage{amsmath,amssymb,amsfonts}
\usepackage{algorithmic}
\usepackage{graphicx}
\usepackage{textcomp}
\usepackage{xcolor}
\usepackage{units}
\usepackage{subcaption}
\usepackage{microtype}
\usepackage{paralist}
\usepackage{multirow}

\usepackage{calrsfs}
\DeclareMathAlphabet{\pazocal}{OMS}{zplm}{m}{n}

\newcommand{\amy}[1]{\textcolor{black}{#1}}

\def\BibTeX{{\rm B\kern-.05em{\sc i\kern-.025em b}\kern-.08em
    T\kern-.1667em\lower.7ex\hbox{E}\kern-.125emX}}
\begin{document}

\title{Feature-informed Embedding Space Regularization For Audio Classification}

\author{\IEEEauthorblockN{Yun-Ning Hung$^1$\thanks{$^1$ This work was performed while Y.~Hung was a master's student at Georgia Institute of Technology.}}
\IEEEauthorblockA{\textit{Speech, Audio \& Music Intelligence Team} \\
\textit{ByteDance}\\
yunning.hung@bytedance.com}
\and
\IEEEauthorblockN{Alexander Lerch}
\IEEEauthorblockA{\textit{Music Informatics Group} \\
\textit{Georgia Institute of Technology}\\
alexander.lerch@gatech.edu}
}

\maketitle

\begin{abstract}
Feature representations derived from models pre-trained on large-scale datasets have shown their generalizability on a variety of audio analysis tasks.
Despite this generalizability, however, task-specific features can outperform if sufficient training data is available, as specific task-relevant properties can be learned. Furthermore, the complex pre-trained models bring considerable computational burdens during inference. 
We propose to leverage both detailed task-specific features from spectrogram input and generic pre-trained features by introducing two regularization methods that integrate the information of both feature classes. The workload is kept low during inference as the pre-trained features are only necessary for training. In experiments with the pre-trained features VGGish, OpenL3, and a combination of both, we show that the proposed methods not only outperform baseline methods, but also can improve state-of-the-art models on several audio classification tasks. The results also suggest that using the mixture of features performs better than using individual features.
\end{abstract}


\section{Introduction}
Feature representations learned with deep architectures from large amounts of training data have shown to be powerful for a variety of tasks beyond the tasks they were trained for. The pre-trained features from the BERT model \cite{devlin2018bert}, for example, have been successfully used in multiple Natural Language Processing (NLP) tasks such as question answering, text classification, and language inference \cite{falke2019ranking}. In the image domain, large-scale pre-trained models such as ImageNet \cite{krizhevsky2012imagenet} and VGG \cite{simonyan2014very} have achieved competitive results in image classification and assist in many visual feature extraction tasks \cite{long2015fully}. 

There exist several pre-trained models in the audio domain, such as L$^3$-Net \cite{cramer2019look}, VGGish \cite{45611}, and Jukebox \cite{dhariwal2020jukebox}. L$^3$-Net leverages both audio and visual information provided by video to train an audio feature extraction network, while the VGGish model is pre-trained on a large-scale audio dataset, AudioSet \cite{45611}. Jukebox adopts a language model similar to NLP and is pre-trained on music audio only. Those features have been successfully applied to various Music Information Retrieval (MIR) tasks, including weakly-supervised instrument recognition \cite{gururani2019attention}, cross-modal representation learning \cite{yu2019deep, zeng2018audio}, music auto-tagging \cite{9439825, castellon2021codified}, music emotion recognition  \cite{koh2021comparison, amiriparian2019emotion}, and music genre classification \cite{ramirez2020machine, huang2020large}. This successful application to a multitude of different tasks combined with the comparably compact representation implies that these features are able to capture many task-agnostic properties of audio and music signals.

Although these transfer learning approaches are used successfully for such a wide range of tasks, there are potential drawbacks when using pre-trained features directly as the input representation. First, the extracted compact representation might lack task-specific information, which, in contrast, a task-specific representation learned from spectrogram input will include \cite{tompkinsdcase}. While it is possible to update and fine-tune the pre-trained model with task-specific datasets to catch this relevant information \cite{xu2018aalto}, the computational burden during training is increased due to the high complexity of the feature extractor. Second, the time resolution of pre-trained features is fixed, forcing any system utilizing the features to the same time resolution. VGGish features, for example, have a time resolution of approx.\ \unit[1]{s}. Tasks that need a higher resolution such as beat-tracking \cite{davies2009evaluation}, might not be able to utilize pre-trained feature representations. Third, extractors such as VGGish and L$^3$-Net rely on fixed input representations such as Mel spectrograms to extract features. Several recent studies, however, have shown that other input representations such as harmonic representations \cite{won2020data} or multi-rate PCEN \cite{ick2021sound} achieve superior performance on several audio-related tasks. Fourth, the deep architectures of pre-trained feature extractors also increase the computational workload and the execution time at inference. Small real-time devices with limited computational power, for example, might not be able to benefit from the pre-trained representations.

To address these drawbacks, we propose to incorporate the task-agnostic information of pre-trained representations into a new smaller-scale model without using them as input to the model. Our approaches show some similarity to feature-based knowledge distillation approaches used in teacher-student learning \cite{gou2021knowledge}, where the pre-trained representations are used to regularize the embedding space during training. Thus, the pre-trained representations provide condensed, task-agnostic information to help shape the embedding, but are not required during inference. The architecture of the original model does not need to change, so that the input representation and its time resolution can be chosen freely for the task at hand. Such feature-based knowledge transfer approaches are rarely explored in the music/audio domain. \amy{Moreover, to extend commonly-used regularization loss functions (e.g., $\pazocal{L}_1$, $\pazocal{L}_\mathrm{CE}$), we propose two loss functions based on cosine distance to match the embeddings. Cosine distance has been commonly used in several metric learning scenarios, such as few-shot learning \cite{vinyals2016matching,gidaris2018dynamic} and recommendation system \cite{hansen2020contextual}.}


To summarize, the main contributions of this paper are:
\begin{inparaenum}[(i)]%
    \item the introduction of two regularization methods integrating the information of pre-trained feature representations during training time without increasing the computational complexity during inference, and
    \item an investigation of the suitability of the two representations VGGish and OpenL3 and their combination for this regularization, and a detailed study of the proposed regularization methods for SOTA models.
\end{inparaenum}

\begin{figure}[t]
  \centering
  \includegraphics[width=\columnwidth]{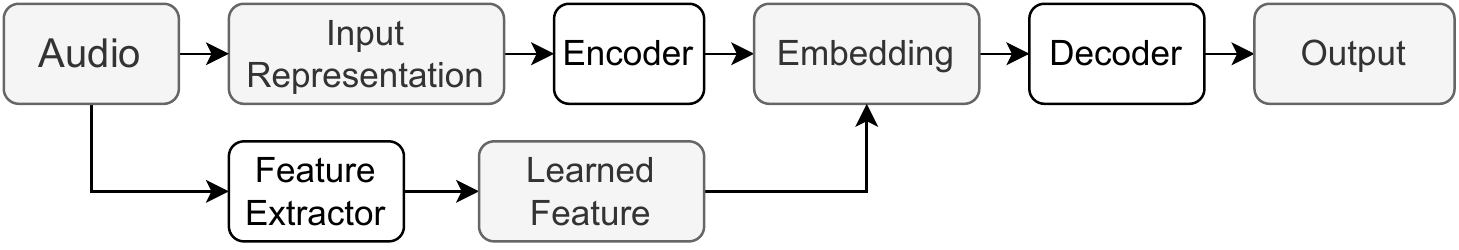}
  \caption{Pipeline for the training including the learned feature. The lower branch is discarded during inference.}
  \label{fig:pipeline}
\end{figure}

\section{Methods}
We propose two approaches to incorporate pre-trained feature representations into existing model architectures. In the following sections, we first introduce the general training pipeline of our regularization methods and the two pre-trained representations we use in our experiment. Then, details of two proposed regularization methods will be presented.

\subsection{System Description}
Figure~\ref{fig:pipeline} shows the general structure of our training pipeline. For the regularization to be applicable, we assume the deep-learning model to be composed of a similar pipeline with an encoder ($E$) to extract the intermediate representation ($l$) and a decoder ($D$) to predict the output based on this representation. The intermediate representation $l \in R^{T_\mathrm{E}\times B_\mathrm{E}}$ is also referred to as embedding, where $T_\mathrm{E}$ represents the number of time frames while $B_\mathrm{E}$ represents the dimensionality of the embedding. 

The pre-trained features are a 2-dimensional representation $v \in R^{T_\mathrm{F}\times B_\mathrm{F}}$ where $T_\mathrm{F}$ represents the number of time frames while $B_\mathrm{F}$
represents the dimensionality of the learned features. Although we only investigate two widely used representations extracted with deep architectures, it is conceivable that any kind of custom feature in this format can be used.

\subsection{Learned Features}
We investigate three regularization inputs in this study, VGGish \cite{45611}, OpenL3 \cite{cramer2019look}, and the combination of both. VGGish and OpenL3 features are chosen due to their proven success in several audio-related downstream tasks \cite{gururani2019attention, zeng2018audio, 9439825, ramirez2020machine}. Both representations are extracted by models trained on large-scale audio datasets. VGGish features are extracted by the VGGish model~\cite{45611} pre-trained on AudioSet \cite{gemmeke2017audio} to perform audio classification at a time resolution of approx.~\unit[1]{s} (no overlap) with a feature dimensionality of $B_\mathrm{F}=128$. The features are PCA transformed (with whitening) and quantized to 8-bits. The OpenL3 features \cite{cramer2019look} are extracted from L$^3$-net pre-trained on a subset of AudioSet. These features are also 2-dimensional representations with a time resolution of \unit[0.1]{s} (no overlap) and $B_\mathrm{F}=512$. 

The combination of these two features is achieved by a simple concatenation along the $B_\mathrm{F}$ dimension. To match the temporal resolution of OpenL3, VGGish features are repeated 10 times along the temporal dimension. The resulting combined representation has a time resolution of \unit[0.1]{s} and a dimensionality of $B_\mathrm{F}=640$.\footnote{More sophisticated interpolation methods for upsampling to match the time resolutions will be explored in future work.} 

\begin{table*}[]
    \centering
    \begin{tabular}{l l c c c c c c c c c}
    \toprule
       & Methods & \multicolumn{4}{c}{DCASE 17 (F1)} & \multicolumn{4}{c}{MTAT (PR-AUC)}\\
       \midrule
       
       & & None & VGGish & OpenL3 & Combined & None & VGGish & OpenL3 & Combined \\
       \cmidrule(r){3-6} \cmidrule{7-10} 
       
       \multirow{4}{*}{Baseline} & Won et al. \cite{won2020data}  &  0.547 & - & - & - & 0.465 & - & - & - \\
       
       & Feature-only & - & 0.496	& 0.477	& 0.501 & - & 0.454 & 0.454	& 0.456 \\
       
       & Concat & - & 0.529 & 0.492 & 0.495 & - & 0.457 & 0.464 &	0.458\\
       
       & FiLM & - & \underline{0.548} & 0.511 & 0.542 & - & \underline{0.470} & 0.463 & \underline{\textbf{0.470}}\\
       
       \cmidrule(r){1-10} 
       
       \multirow{2}{*}{Proposed} & Con-Reg  & - & \underline{\textbf{0.568}} & \underline{\textbf{0.557}} & \underline{\textbf{0.576}} & - & \underline{\textbf{0.471}} & \underline{0.466} & \underline{0.469} \\
       
       & Dis-Reg & - & \underline{0.548} & 0.543 & \underline{0.563} & - & 0.464 & \underline{\textbf{0.468}} & 0.463 \\
      
    \bottomrule

    \end{tabular}
    \caption{Evaluation results for the DCASE 17 and MTAT datasets. Bold numbers represent the highest score in each category while the underlined numbers indicate the scores improved over the baseline model.} 
    \label{tab:result}
\end{table*}

\subsection{Proposed Feature Integration Methods}

Since VGGish, OpenL3, and the mixture features have different time resolution than the embeddings from most of the existing models, we repeat or average-pool the features $n$ times with $n$ representing the number of time frames in \unit[1]{s} or \unit[0.1]{s} to fit the time resolution. This approach is based on the assumption that a slight misalignment in time will not influence the result too much.

The design of the regularization methods is based on two assumptions. First, pre-trained features might contain information that is useful for various tasks but cannot be adequately represented in the unregularized embedding space (e.g., due to insufficient training data). Second, pre-trained features have strong discriminative power. The proposed regularization methods attempt to transfer the knowledge from the pre-trained feature space into the embedding vectors by adding structure to the embedding vectors resulting in a more separable embedding space.

Both methods add an extra loss term for network training:
\begin{equation}
      L_\mathrm{final} = L + \alpha L_\mathrm{Reg} ,
\end{equation}
with the hyperparameter $\alpha$ adjusting the contribution on the overall loss $L_\mathrm{final}$.
The following methods are proposed to incorporate learned features into the embeddings:

\begin{itemize}
    
    \item \textbf{\textit{Con-Reg}}: {Con-Reg}  aims at regularizing the embedding space so that its layout becomes more similar to the feature space of the pre-trained feature $v$. To do so, we utilize the features extracted from the audio and add an extra loss term $L_\mathrm{Con-Reg}$ based on cosine distance\footnote{Pilot experiments using $\pazocal{L}_2$ as the distance function did not lead to competitive results.} to minimize the distance between embeddings and pre-trained features: 
    \begin{equation}\label{eq:vgg-close}
      L_\mathrm{Con-Reg} = d_\mathrm{cos}(l, f(v)) 
    \end{equation}
    where $f$ represents a 1D CNN with kernel size \unit[1] \cite{szegedy2015going}
    to transform the feature dimensionality $B_\mathrm{F}$ to match the embedding dimensionality, allowing us to compute the cosine distance between $l$ and $v$. 
    
    \item \textbf{\textit{Dis-Reg}}: Dis-Reg is a distance-based regularization. Similar to \textit{Con-Reg}, the embedding space is regularized with an additional loss. In this case, however, the additional loss term $L_\mathrm{Dis-Reg}$ aims at forcing the distances between pairs of embedding vectors to be similar to the distance of corresponding pairs of pre-trained features:
    \begin{equation}
      L_\mathrm{Dis-Reg} = |d_\mathrm{cos}(l_i, l_j) - d_\mathrm{cos}(v_i, v_j)| ,
    \end{equation}
    where $d_\mathrm{cos}$ represents the cosine distance between two embeddings from samples $i$ and $j$ ($i \neq j$) or the distance of two corresponding learned features, respectively.
\end{itemize}




\subsection{Model Architecture}
The harmonic CNN model proposed by Won et al.\ \cite{won2020data} is chosen as our experimental system. This model utilizes learnable harmonic filters to capture the inherent harmonic structure of the input audio and achieves SOTA results on music tagging, sound event tagging, and keyword spotting. 
The model is composed of seven residual blocks as an encoder to extract embeddings from the harmonic representation. Two linear layers are then used as a decoder to predict the output from the embeddings.  

\section{Experiment} \label{sec:exp}

\subsection{Baseline Feature Integration Methods}
To provide a baseline reference as a comparison to the proposed regularization methods, we also present the result for three baseline methods incorporating pre-trained features:
\begin{itemize}
    \item \textbf{\textit{Features-only}}: The simplest way of incorporating pre-trained features is to directly use these pre-trained features as the input for the decoder. The decoder of the original model is adjusted to fit the dimensionality of the features. 
    \item \textbf{\textit{Concat}}: Concat simply concatenates embeddings $l$ and pre-trained features $v$ along the $B$ dimension. The pre-trained features, therefore, act as a supplementary of the embeddings and the embedding space is transformed into a joint feature space. The decoder then has the flexibility to decide how to leverage the information from the joint feature space. Since the concatenated embeddings have a larger dimension along $B$ than the original embeddings, the first layer of the decoder will be adjusted to fit the dimensionality of the new intermediate representation. 
    
    \item \textbf{\textit{FiLM}}: A Feature-wise Linear Modulation (FiLM) layer is originally proposed as a general-purpose conditioning method to assist visual reasoning, which is difficult for standard deep-learning methods \cite{perez2018film}. The success of FiLM leads to its successful usage in several audio-related tasks \cite{slizovskaia2019end, kim2019neural}. The FiLM layer influences neural network computation via a simple, feature-wise affine transformation based on conditioning information:
    \begin{equation}
    l_{new} = \gamma\cdot l + \beta
    \end{equation}
    where $\gamma=f(v)$ and $\beta=h(v)$ are calculated based on pre-trained features $v$. We use a simple linear layer for the functions $f$ and $h$ to compute the transformation parameters $\gamma$ and $\beta$. The resulting embeddings after the transformation are $l_{new}$. 
\end{itemize}

\subsection{Tasks}
We investigate our proposed methods on two different tasks to demonstrate the influence of the regularization: music tagging and sound event tagging as also used by Won et al.\ \cite{won2020data}. 

\subsubsection{Music Auto-Tagging}
Music auto-tagging is a multi-label classification problem aiming to predict tags for a musical piece. The used dataset is MagnaTagATune (MTAT) \cite{law2009evaluation}. This dataset has approx.~21k audio clips with each clip around \unit[30]{s}. The dataset contains a variety of tags, including genre, mood, and instrumentation. The top-50 tags are chosen for label prediction. The results are reported using the Area Under Precision-Recall Curve (PR-AUC). The OpenL3 features are extracted in the music subset mode. 

\subsubsection{Sound Event Tagging}
Sound event tagging is an important task that has been included in DCASE challenges for many years \cite{mesaros2017dcase}. The goal of this task is to detect audio events in a sound excerpt. Following Won et al., we choose ``Task 4: Large-scale weakly supervised sound event detection for smart cars'' from the DCASE2017 challenge as our target task. The dataset is a subset of AudioSet \cite{gemmeke2017audio} and contains approx.~53k audio excerpts for 17 classes. Each excerpt is around \unit[10]{s}. The results are reported using the average of instance-level F1-scores with a threshold value of 0.1. In this case, the OpenL3 features are extracted in the environmental subset mode. 

\subsection{Experimental Setup}
The training setup, including dataset cleaning and split and the selection of optimizer and learning rate, is mirrored from their work. All methods are evaluated on each task by using VGGish, OpenL3, and the combined pre-trained features. The best model is
selected based on the validation metric for each task for testing. After hyperparameter tuning, we choose $\alpha=5$ for $L_\mathrm{Con-Reg}$ and $\alpha=1$ for $L_\mathrm{Dis-Reg}$.

\begin{figure}[t]
  \centering
  \includegraphics[width=\columnwidth]{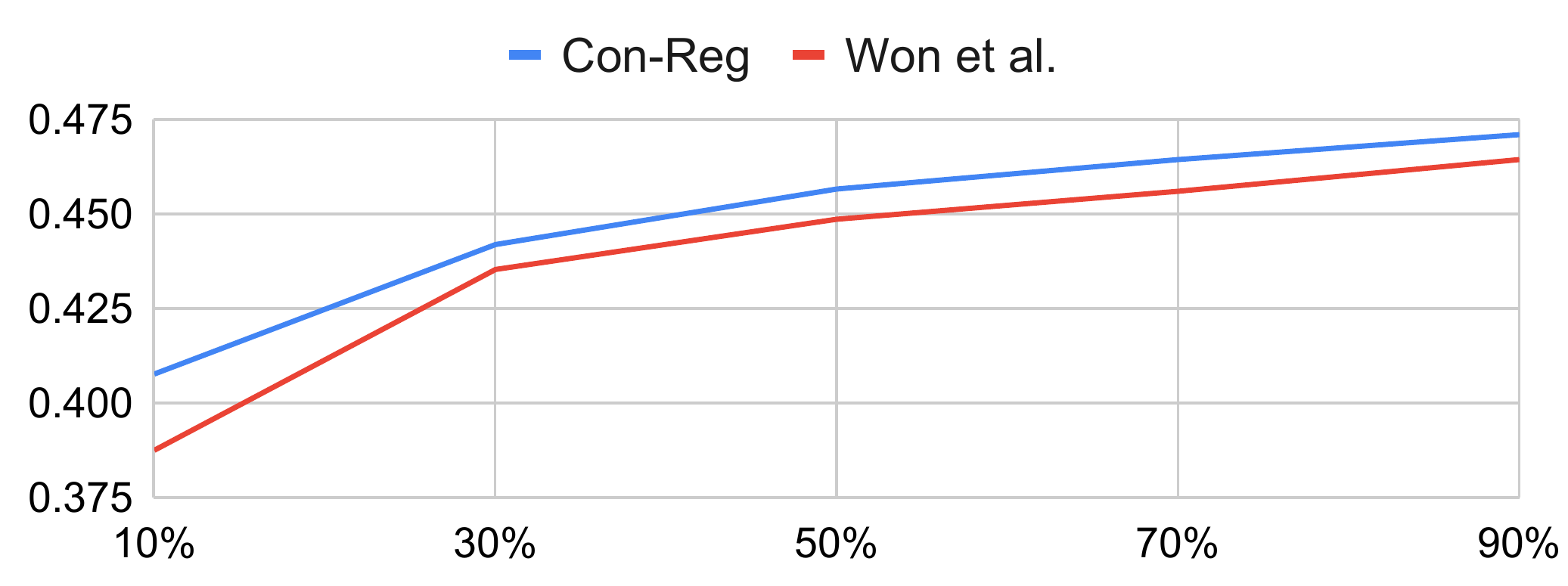}
  \caption{Ablation study for the results under different percentage of training data for SOTA system and our best method.}
  \label{fig:ablation}
\end{figure}

\section{Result and Discussion}
The experimental results are shown in Table~\ref{tab:result}. In the following, we discuss different aspects of these results separately

\subsubsection{Learned Features}
\label{sec: feature}
It can be observed that VGGish features generally outperform the OpenL3 features. The combined features tend to perform on par or better than the individual features. This result suggests that ---although VGGish and OpenL3 were both trained on AudioSet--- they encode somewhat complementary information. This is most likely due to different training strategies and hints at the possibility of combining them for other downstream tasks.  

\subsubsection{Baselines}
Among all the baseline methods, \emph{FiLM} and \emph{Concat} generally perform better than using only pre-trained features for prediction, but comparable to or only slightly better than Won et al.'s original system. This result indicates that there is some task-specific information that exists in the harmonic representations but not in the pre-trained features. \emph{FiLM}, in general, tends to outperform \emph{Concat}.

\subsubsection{Proposed Methods}
Both proposed methods tend to generally give higher results than the baselines and thus demonstrate the efficiency of our proposed regularization methods. Since the pre-trained features could not achieve better performance than learning from harmonic representation, directly incorporating features into embedding can potentially be seen as noise during inference and drop the performance. In contrast, indirectly leveraging pre-trained features does not have this limitation. The model is able to combine the task-specific information learned through the harmonic representation with the complementary, task-agnostic but highly meaningful information of the pre-trained features.

\emph{Con-Reg} generally outperforms \emph{Dis-Reg}. We discover that  $L_\mathrm{Dis-Reg}$ has larger values than $L_\mathrm{Con-Reg}$ during training. The larger loss term indicates that optimizing the model to predict embedding vectors matching the distance of pre-trained features is harder than directly minimizing the distance between embedding vectors and pre-trained features. As a result, this regularization method could not structure embeddings space better than \textit{Con-Reg}.

Furthermore, \emph{Con-Reg} achieves better results than the SOTA model proposed by Won et al.\ \cite{won2020data} on both the DCASE 17 dataset and MTAT dataset.

\subsubsection{Amount of Training Data}
We can observe that the improvement for the MTAT dataset is very limited compared to the DCASE 17 dataset. This might be due to the fact that the MTAT dataset is larger in size; therefore, the model is able to learn good embeddings directly from the training data and the additional information from the learned features is unnecessary. To compare the effect of insufficient training data, we present the results of an ablation study by randomly choosing $x$ percent of audio files in the training set to form training subsets with $x \in {10\%, 30\%, 50\%, 70\%, 90\%}$. We use the best setting, \textit{Con-Reg} with combined features, for the experiment and compare the performance with the SOTA system. The results are shown in Figure~\ref{fig:ablation}. We can observe that with decreasing training data, the improvement from the proposed regularization method becomes more obvious, especially lower than 30\% of training data.


\begin{table}[]
    \centering
    \begin{tabular*}{\columnwidth}{l c c c c c c}
    \toprule
       Methods & \multicolumn{3}{c}{Training Parameters} & \multicolumn{3}{c}{Inference Parameters} \\
       \midrule
        & VGG & L3 & Comb. & VGG & L3 & Comb. \\
       \cmidrule(r){2-4} \cmidrule{5-7}
       Feature-only\phantom{...} & 38k	& 136k & 169k & 72.2M & 4.8M & 77.0M \\
       Concat & 3.7M & 3.8M & 3.8M & 75.8M & 8.4M & 80.6M \\
       FiLM & 3.7M & 3.9M & 4.0M & 75.8M & 8.6M & 80.8M \\
       Con-Reg & 3.7M & 3.8M & 3.8M & 3.6M & 3.6M & 3.6M\\
       Dis-Reg & 3.6M & 3.6M & 3.6M & 3.6M & 3.6M & 3.6M \\
    \bottomrule

    \end{tabular*}
    \caption{Number of training and testing parameters for both proposed and baseline methods. }
    \label{tab:parameters}
\end{table}

\subsection{Model Complexity}

Table~\ref{tab:parameters} shows the number of training and inference parameters for both the proposed and the baseline methods. 
Since audio files can be pre-processed to extract the learned features, the \emph{feature-only} method only needs the decoder during training and has, therefore, the smallest number of trainable parameters. In contrast, the same \emph{feature-only} method requires a large number of parameters during inference due to the large-scale VGGish/OpenL3 extractor. A similar imbalance between the number of training and inference parameters can be observed for the \emph{Concat} and \emph{FiLM} methods.

Comparing the baseline results from Table~\ref{tab:result} with the number of baseline parameters listed in Table~\ref{tab:parameters} we can conclude that the better the result, the more parameters the extractor needs. For example, VGGish features achieve better performance than OpenL3 but its extractor is also more complex. As mentioned above, the combination of pre-trained features can achieve better performance, but the parameters for inference increase considerably as they require two extractors (both VGGish and OpenL3) during inference.

The proposed regularization methods have a similar number of parameters as the original model proposed by Won et al.~\cite{won2020data} (approx.~3.6M), for both training and inference. This is because our proposed methods only use features in the additional regularization loss function without modifying the model architecture.

\section{Conclusion}
In this work, we proposed two novel regularization methods to incorporate the information from pre-trained features during training by adding an additional loss term restructuring the embedding space. 
The proposed regularization methods show improvements compared to baseline feature integration methods, which either directly use pre-trained features or directly influence the model through concatenation or a feature-wise linear transform. Furthermore, the regularized models can outperform SOTA audio classification models, especially having a more pronounced performance increase in the case of limited training data.

Future work will include experiments on other features, such as Jukebox \cite{dhariwal2020jukebox} and Musicnn \cite{pons2019musicnn}, as well as the exploration of different interpolation methods to deal with the various time resolutions. We also plan to extend our experiments to non-audio-related fields such as NLP and computer vision.


\bibliographystyle{IEEEtran}
\bibliography{IEEEtran}

\end{document}